\documentclass[aps,prb,reprint,superscriptaddress,longbibliography,floatfix]{revtex4-2}

\usepackage{amsmath,amssymb,bm}
\usepackage{graphicx}
\usepackage{booktabs}
\usepackage{xcolor}
\usepackage[colorlinks=true,allcolors=blue]{hyperref}

\graphicspath{{figures/}}
\hypersetup{
  pdftitle={Broadband phonon-velocity suppression and a finite anisotropic crossover in twisted bilayer SnSe},
  pdfauthor={Peng Kang et al.}
}

\begin{document}

\title{Broadband phonon-velocity suppression and a finite anisotropic crossover in twisted bilayer SnSe}

\author{Peng Kang}
\affiliation{School of Materials Science and Engineering, Beihang University,
No. 37 Xueyuan Road, Beijing 100191, China}
\affiliation{State Key Laboratory of Artificial Intelligence for Material
Science, Beihang University, No. 37 Xueyuan Road, Beijing 100191, China}
\affiliation{Tianmushan Laboratory, Beihang University, Hangzhou 311115, China}

\author{Wei Yin}
\affiliation{Tianmushan Laboratory, Beihang University, Hangzhou 311115, China}

\author{Da Wan}
\affiliation{School of Materials Science and Engineering, Beihang University,
No. 37 Xueyuan Road, Beijing 100191, China}
\affiliation{State Key Laboratory of Artificial Intelligence for Material
Science, Beihang University, No. 37 Xueyuan Road, Beijing 100191, China}
\affiliation{Tianmushan Laboratory, Beihang University, Hangzhou 311115, China}

\author{Shulin Bai}
\affiliation{School of Materials Science and Engineering, Beihang University,
No. 37 Xueyuan Road, Beijing 100191, China}
\affiliation{State Key Laboratory of Artificial Intelligence for Material
Science, Beihang University, No. 37 Xueyuan Road, Beijing 100191, China}
\affiliation{Tianmushan Laboratory, Beihang University, Hangzhou 311115, China}
\affiliation{Center for Bioinspired Science and Technology, Hangzhou
International Innovation Institute, Beihang University, Hangzhou 311115,
China}

\author{Sirui Fan}
\affiliation{School of Materials Science and Engineering, Beihang University,
No. 37 Xueyuan Road, Beijing 100191, China}
\affiliation{State Key Laboratory of Artificial Intelligence for Material
Science, Beihang University, No. 37 Xueyuan Road, Beijing 100191, China}
\affiliation{Tianmushan Laboratory, Beihang University, Hangzhou 311115, China}

\author{Qi Zou}
\affiliation{Tianmushan Laboratory, Beihang University, Hangzhou 311115, China}

\author{Hongfeng Li}
\affiliation{School of Materials Science and Engineering, Beihang University,
No. 37 Xueyuan Road, Beijing 100191, China}
\affiliation{State Key Laboratory of Artificial Intelligence for Material
Science, Beihang University, No. 37 Xueyuan Road, Beijing 100191, China}
\affiliation{Tianmushan Laboratory, Beihang University, Hangzhou 311115, China}

\author{Xiao Xiang}
\affiliation{School of Materials Science and Engineering, Beihang University,
No. 37 Xueyuan Road, Beijing 100191, China}
\affiliation{State Key Laboratory of Artificial Intelligence for Material
Science, Beihang University, No. 37 Xueyuan Road, Beijing 100191, China}
\affiliation{Tianmushan Laboratory, Beihang University, Hangzhou 311115, China}

\author{Zhen Li}
\affiliation{School of Materials Science and Engineering, Beihang University,
No. 37 Xueyuan Road, Beijing 100191, China}
\affiliation{State Key Laboratory of Artificial Intelligence for Material
Science, Beihang University, No. 37 Xueyuan Road, Beijing 100191, China}
\affiliation{Tianmushan Laboratory, Beihang University, Hangzhou 311115, China}

\author{Yu Liu}
\affiliation{School of Materials Science and Engineering, Beihang University,
No. 37 Xueyuan Road, Beijing 100191, China}
\affiliation{State Key Laboratory of Artificial Intelligence for Material
Science, Beihang University, No. 37 Xueyuan Road, Beijing 100191, China}
\affiliation{Tianmushan Laboratory, Beihang University, Hangzhou 311115, China}

\author{Lei Zheng}
\email{zhenglei@buaa.edu.cn}
\affiliation{School of Materials Science and Engineering, Beihang University,
No. 37 Xueyuan Road, Beijing 100191, China}
\affiliation{State Key Laboratory of Artificial Intelligence for Material
Science, Beihang University, No. 37 Xueyuan Road, Beijing 100191, China}
\affiliation{Tianmushan Laboratory, Beihang University, Hangzhou 311115, China}

\author{Li-Dong Zhao}
\email{zhaolidong@buaa.edu.cn}
\affiliation{School of Materials Science and Engineering, Beihang University,
No. 37 Xueyuan Road, Beijing 100191, China}
\affiliation{State Key Laboratory of Artificial Intelligence for Material
Science, Beihang University, No. 37 Xueyuan Road, Beijing 100191, China}
\affiliation{Tianmushan Laboratory, Beihang University, Hangzhou 311115, China}
\affiliation{Center for Bioinspired Science and Technology, Hangzhou
International Innovation Institute, Beihang University, Hangzhou 311115,
China}
\date{\today}

\begin{abstract}
Moir\'e superlattices reshape lattice dynamics without altering chemical
composition, yet how crystal anisotropy modifies this control remains
unclear.  We combine density-functional-theory (DFT)-calibrated
lattice-dynamical calculations with angle-matched untwisted controls to study
puckered bilayer SnSe across seven commensurate twist angles
($3.18^\circ$--$8.77^\circ$).  At 300 K, twisting suppresses the band-path
heat-capacity-weighted mean-square group velocity to 2.6--8.4\% of the
control values; the suppression spans a broad frequency range rather than a
few soft branches.  The velocity response crosses over between $4.78^\circ$
and $3.82^\circ$ into a regime where the relaxed stacking textures and
frequency-resolved velocity profiles become self-similar, with the normalized
mean-square velocity ratio spanning only 11.1\% of its mean across the three
smallest angles---a finite anisotropic crossover, not a singular-angle
condition.  Direct DFT--MACE force-constant agreement ($r=0.996$), uniform
$4\times4\times1$ stability scans, and acoustic-sum-rule and path-density
tests support the trend.  The equilibrium trend is defined by six structures
after excluding one relaxation-sensitive case.  These results extend phonon
twistronics to low-symmetry layered materials and identify crystal anisotropy
as a key determinant of finite-angle phonon crossover behavior.
\end{abstract}

\maketitle

\begin{figure*}[t!]
  \centering
  \includegraphics[width=\textwidth]{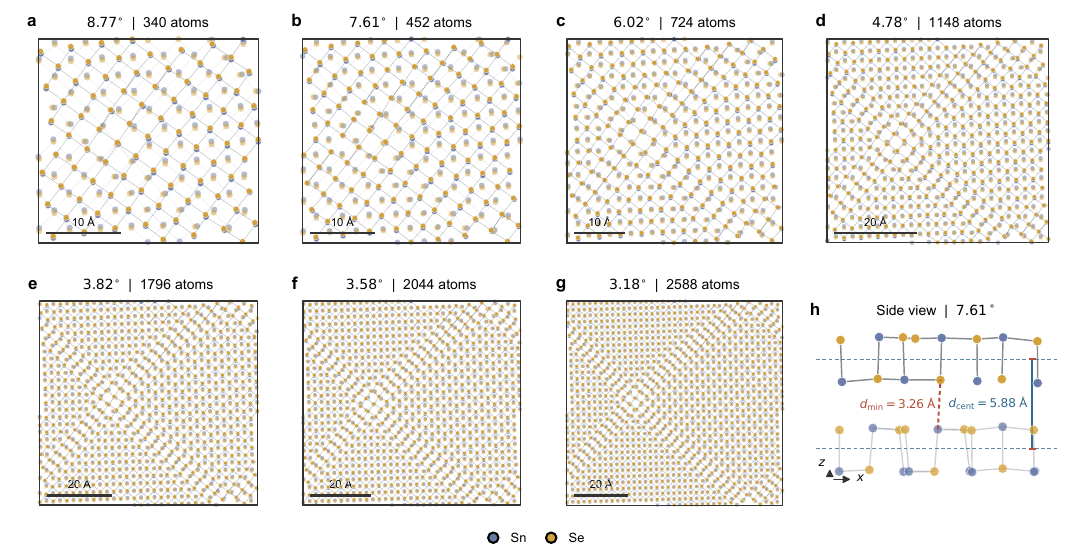}
  \caption{\label{fig:structure}Relaxed vacuum-separated twisted bilayer SnSe.
  Top views (a)--(g) are ordered by decreasing twist angle and increasing atom
  count: $8.77^\circ$, $7.61^\circ$, $6.02^\circ$, $4.78^\circ$,
  $3.82^\circ$, $3.58^\circ$, and $3.18^\circ$.  (h) Representative side view
  of the $7.61^\circ$ cell, showing the puckered layers, layer-center
  separation $d_{\rm cent}$, and minimum interlayer distance $d_{\rm min}$.
  Complete cell dimensions and periodic vacuum gaps are listed in
  Table~\ref{tab:cells}.  Dark blue and gold denote Sn and Se, respectively.}
\end{figure*}

\begin{figure}[t!]
  \centering
  \includegraphics[width=\columnwidth]{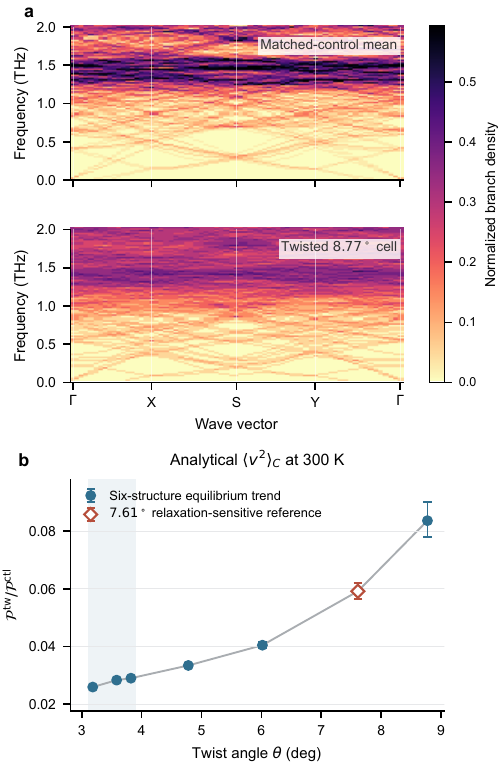}
  \caption{\label{fig:phonon}Direct spectral and angle-dependent evidence for
  harmonic velocity suppression.  (a) Low-frequency folded-branch density along
  $\Gamma$--$X$--$S$--$Y$--$\Gamma$ for the $8.77^\circ$ twisted cell and the
  mean of its two matched controls, shown on identical scales.  The branch
  density is divided by the number of branches and the frequency-bin width, so
  its frequency integral is unity at each path point; the twisted spectrum
  redistributes the sharp control ridges into broader, less dispersive folded
  features.  (b) The 300-K
  analytical $\mathcal{P}=\langle v^2\rangle_C$ ratio
  [Eq.~(\ref{eq:proxy})] relative to the
  per-angle matched-control mean.  Error bars span the two individual-control
  normalizations.  The $7.61^\circ$ structure is marked separately because
  its original $S$-point feature disappears after tighter fixed-cell
  relaxation; it is excluded from the six-structure equilibrium trend.  The
  gray band marks the three smallest-angle structures,
  $3.18^\circ$--$3.82^\circ$.  Lines connect calculated points as a visual
  guide.}
\end{figure}

\section{Introduction}

Controlling heat-carrying phonons without degrading electronic functionality
is a central challenge in thermoelectric materials design.  Composition,
defects, bonding, strain, and data-guided optimization provide established
routes to this goal \cite{Snyder2008,Bai2026Thermoelectrics,Wan2025Mg}, while
lone-pair activity and higher-order scattering show how sensitively lattice
dynamics can respond to local structure in layered chalcogenides
\cite{Wan2025BiCuSeO,Bai2025SrSnSe2}.  Puckered SnSe is a particularly
revealing platform: its low thermal conductivity and strong in-plane
anisotropy underpin high thermoelectric performance
\cite{Zhao2014,Zhao2016,Qin2024}, and its soft bonding, low-frequency modes,
and anharmonicity produce an unusually responsive phonon spectrum
\cite{Li2015,Bansal2016,Guo2015}.  A structural control variable that leaves
the chemical composition unchanged is therefore especially attractive.

Interlayer twist supplies such a variable by creating a spatially varying
stacking registry.  The resulting moir\'e superlattice reconstructs, folds
phonon branches into a reduced Brillouin zone, and can reorganize shear,
breathing, acoustic, and low-energy optical modes
\cite{Cocemasov2013,Koshino2019,Lu2022,Samajdar2022,Girotto2023,Lin2018}.
Flat optical branches and tunable moir\'e phonons have been identified in
graphene and transition-metal dichalcogenide bilayers
\cite{Cappelluti2023,RamosAlonso2025}; at small angles, relaxation can produce
pinned domains, soliton networks, and spatially localized lattice dynamics
\cite{Ochoa2019,Maity2020,Gadelha2021}.  Reduced thermal conductivity has also
been reported in hexagonal SnSe$_2$ moir\'e superlattices and incommensurate
chalcogenide stacks \cite{Ran2025,Xu2025}.  These advances establish twist as a
phonon-control mechanism, but they primarily concern high-symmetry lattices or
selected low-energy branches.

Low-symmetry SnSe poses a distinct scientific question.  Its two in-plane axes
are inequivalent, so both elastic accommodation and registry forces are
direction dependent.  The associated reconstruction parameters need not
become nonlinear at a single angle, suggesting a finite anisotropic crossover
rather than a singular-angle condition.  It is not known whether this reconstruction
produces only isolated soft branches or a broadband change in phonon
propagation, nor whether the structural texture and spectral response approach
a common small-angle form.  Resolving these alternatives requires
matched untwisted controls: otherwise changing source-layer metrics,
moir\'e cell size, and branch multiplicity can be mistaken for an intrinsic
twist effect.  The relevant observable must also separate group-velocity
renormalization from thermal occupation without being conflated with absolute
lattice thermal conductivity \cite{Strongman2021,Feng2023}.

The required calculations span hundreds to thousands of atoms and are beyond
routine DFT finite displacements.  Machine-learned interatomic potentials make
this scale accessible, but phonons depend on local force derivatives rather
than energies alone \cite{Rowe2018}.  A credible large-cell analysis therefore
needs direct DFT validation of harmonic force constants, isolated bilayers at
physical separations, matched structural references, analytical group
velocities, and uniform q-grid stability checks.  These requirements are especially
important in dense folded spectra, where branch tracking and shallow numerical
modes can otherwise obscure the physical trend.

Here we combine a DFT-calibrated MACE potential
\cite{Batatia2022,Batatia2024} with centered finite-displacement
phonons for seven vacuum-separated commensurate bilayer SnSe cells spanning
$3.18^\circ$--$8.77^\circ$ and 340--2588 atoms.  Each twisted cell is compared
with two registry-minimum controls, whose mean defines the matched control, and
the local harmonic response is tested directly against DFT.  We find a
broadband suppression of the
heat-capacity-weighted phonon velocity scale that strengthens with increasing
moir\'e length, followed by a slowly varying small-angle regime.  The onset
between $4.78^\circ$ and $3.82^\circ$ is identified independently by the
self-similarity of the relaxed stacking texture and the frequency-resolved
velocity response.  An anisotropic elastic--registry framework connects these
observations and establishes interlayer twist as a broader route toward
lattice-dynamical control in low-symmetry moir\'e materials.

The remainder of this paper is organized as follows.
Section~\ref{sec:methods} describes the commensurate-cell construction and
matched controls, the DFT-calibrated MACE potential, the centered
finite-displacement phonon calculations, and the analytical velocity
descriptor.  Section~\ref{sec:results} presents the relaxed moir\'e
structures, the angle-dependent velocity suppression, its frequency-resolved
origin, and the small-angle crossover.  Section~\ref{sec:discussion} develops
the anisotropic elastic--registry scaling framework and defines the scope of
the descriptor.  Section~\ref{sec:conclusions} concludes and outlines
extensions.  Appendices~\ref{app:construction}--\ref{app:methods} collect
structural parameters, registry energetics, the force-constant benchmark,
convergence and stability tests, selected-mode dynamics, and the crossover
correlation analysis.

\section{Methods}
\label{sec:methods}

\subsection{Commensurate bilayers and matched controls}

The rectangular commensurate cells were described by the in-plane integer
matrices
\begin{equation}
 \bm{M}_{\mathrm{L}} =
 \begin{pmatrix}p&0\\0&q\end{pmatrix}, \qquad
 \bm{M}_{\mathrm{U}} =
 \begin{pmatrix}p&s\\-t&q\end{pmatrix},
 \label{eq:commensurate-matrices}
\end{equation}
for the lower and upper layers.  We use $s=t=1$, for which
$\tan^2\theta=1/(pq)$.  The seven pairs $(p,q)=(7,6)$, $(8,7)$, $(10,9)$,
$(13,11)$, $(16,14)$, $(17,15)$, and $(19,17)$ give twist angles of
$8.77^\circ$, $7.61^\circ$, $6.02^\circ$, $4.78^\circ$, $3.82^\circ$,
$3.58^\circ$, and $3.18^\circ$.  The corresponding cells contain 340, 452,
724, 1148, 1796, 2044, and 2588 atoms.  The relaxed structures are shown in
Fig.~\ref{fig:structure}; complete commensuration, strain, cell, and phonon
information is given in Appendix~\ref{app:construction}.

The lower and upper layers were generated from independently audited SnSe
primitive motifs.  Exact rectangular closure was obtained by equal and
opposite logarithmic area accommodation of the two layers, followed by an
area-preserving aspect correction.  The primitive-index-zero Sn sites were
placed at the same in-plane cell center before relaxation; this choice fixes
the otherwise arbitrary origin of the moir\'{e} registry field.  Source-cell
metrics, construction conventions, and strain bounds are given in
Appendix~\ref{app:construction}.

Each cell has a fixed out-of-plane lattice parameter of 33.934 \AA{} and was
initialized with a 25-\AA{} periodic vacuum gap.  The initial layer-center
separation, 5.913 \AA, was the mean of the DFT registry atlas described below.
Atomic
coordinates were relaxed at fixed cell with the calibrated MACE potential and
the FIRE optimizer in ASE 3.27.0 until the maximum force was below
$10^{-3}$ eV \AA$^{-1}$.  A zero-step force evaluation then confirmed the
convergence of each stored structure.  The final periodic vacuum gaps are
24.65--24.93 \AA, so the relaxed bilayers remain isolated from their periodic
images.

For every angle, two untwisted registry-minimum controls were constructed by
duplicating the lower- and upper-source lattices separately.  This preserves
the two layer metrics that form the commensurate twisted cell.  Both controls
were relaxed with the same cell height, interlayer initialization, potential,
and force criterion.  Their arithmetic mean defines the matched control for
each reported phonon quantity.  Registry coordinates depend on the source
primitive origin; the control translation is the stable low-energy registry
equivalent to the DFT minimum, $(u,v)=(1/3,0)$, in the corresponding
source-layer convention.

\subsection{DFT-calibrated MACE potential}

The interatomic potential was initialized from the medium MACE-MPA-0 model
\cite{Batatia2022,Batatia2024} and calibrated with 536 DFT structures sampled
from nine bilayer registries and their relaxation trajectories.  A
registry-grouped split contained 369 training, 118 validation, and 49 test
structures.  On the held-out test set, the energy and force-component
root-mean-square errors (RMSEs) are 0.805 meV atom$^{-1}$ and
0.0141 eV \AA$^{-1}$, respectively.  For the relaxed registry
structures, the relative-energy root-mean-square error is 4.31 meV per
bilayer, the rank correlation is 1.000, and the DFT minimum registry is
reproduced.  The training labels use the PBE-D3(BJ), Sn$_{\mathrm d}$/Se,
600-eV, and $16\times16\times1$ settings specified for the registry atlas
below.

The inherited medium model has two interaction layers, a 6.0-\AA{} radial
cutoff, hidden irreducible representations $128\times0e+128\times1o$,
correlation order 3 (body order 4), and spherical harmonics through
$\ell=3$.  Fine tuning used a \texttt{ScaleShiftMACE} model in 64-bit
precision, estimated elemental reference energies, equal energy and force
loss weights of 100, and configuration weights 1:1:3 for the default and two
registry-trajectory groups.  The embedding and interaction linear weights
were frozen.  Adam with AMSGrad was used for 30 epochs with learning rate
$3\times10^{-4}$, weight decay $5\times10^{-7}$, batch size 16,
exponential-moving-average decay 0.995, and random seed 20260807.  The model
was trained with MACE 0.3.16 and PyTorch 2.10.0.

The fine-tuning set samples discrete registries and their relaxation paths but
does not contain an independent homogeneous-strain series.  The commensurate
cells have at most 0.60\% area accommodation and 0.81\% aspect correction,
and the per-angle controls retain those same metrics.  This construction
cancels the leading metric dependence in the reported ratios; the direct DFT
force-constant comparison below remains the validation of the local harmonic
response rather than a claim of unrestricted transferability.

The local harmonic response was tested separately with centered
$\pm0.01$-\AA{} DFT displacements.  The comparison contains all 24 columns of
the minimum-registry primitive-cell Hessian, selected Sn and Se columns at two
additional registries, and representative responses in a $2\times2$ cell.
The resulting force-constant and phonon comparisons are given in
Appendix~\ref{app:ifc}.  The nine matched optical modes below 2 THz have a
frequency mean absolute error (MAE) of 0.023 THz; thus the benchmark includes the
low-frequency shear, breathing, and optical sector that carries substantial
weight in the harmonic descriptor.

\subsection{Finite-displacement phonons}

Harmonic force constants were generated with Phonopy \cite{Togo2023} from
MACE forces.  Centered positive and negative Cartesian displacements were
applied in each commensurate cell with no further real-space replication.  The
reference displacement amplitude was 0.010 \AA.  No symmetry reduction was
used, so every $N$-atom cell required $6N$ force evaluations.  Across the seven
twisted cells and fourteen matched controls, the calculation contains 163,656
displaced configurations.

The force constants were processed with one successive application of
Phonopy's translational and permutation symmetries.  Thus the production
force constants satisfy the acoustic sum rule (ASR); this operation is more
specific than space-group averaging.  Phonopy 4.1.0, ASE 3.27.0, NumPy 2.1.1,
and MACE 0.3.16 were used for the finite-displacement workflow.  The effect of
using the preserved pre-ASR force constants and positive-frequency cutoffs of
0.02, 0.05, and 0.10 THz is reported in
Appendix~\ref{app:low-frequency}.

Phonon frequencies for band display were evaluated along the common rectangular
path $\Gamma$--$X$--$S$--$Y$--$\Gamma$ with 18 points per segment.  The
band-path minima in Tables~\ref{tab:force} and \ref{tab:low-frequency} use this
display path, whereas the velocity descriptor below uses nine segment
midpoints.
Low-frequency stability was independently scanned on a uniform
$4\times4\times1$ q-grid for every twisted cell using the same ASR-processed
force constants.  Displacement-amplitude and structural-relaxation sensitivity
tests for the relaxation-sensitive $7.61^\circ$ reference, including a tightly
relaxed centered endpoint spectrum, are reported in
Appendix~\ref{app:displacement}.

\subsection{Harmonic velocity descriptor}

For a frequency $\nu$ at temperature $T$, the harmonic mode heat capacity is
\begin{equation}
 \frac{C(\nu,T)}{k_{\mathrm B}} =
 x^2\frac{e^x}{(e^x-1)^2}, \qquad
 x=\frac{h\nu}{k_{\mathrm B}T}.
\end{equation}
Group velocities were evaluated from Phonopy's analytical derivative of the
dynamical matrix,
\begin{equation}
 \bm v_{qj}=\frac{1}{2\omega_{qj}}
 \left\langle \bm e_{qj}\left|
 \frac{\partial \bm D(q)}{\partial \bm q}
 \right|\bm e_{qj}\right\rangle .
 \label{eq:analytic-velocity}
\end{equation}
The Cartesian velocity was projected along each segment of
$\Gamma$--$X$--$S$--$Y$--$\Gamma$ as
$v_{qj}^{\parallel}=\bm v_{qj}\!\cdot\!\hat{\bm t}_s$, where
$\hat{\bm t}_s$ is the unit tangent of segment $s$.  Nine equally spaced
midpoints were used
per segment, and the vertices were excluded from the average.  Because
Eq.~(\ref{eq:analytic-velocity}) is evaluated at each q point from its
eigenvectors and $\partial\bm D/\partial\bm q$, no adjacent-q band-index
matching or finite difference across band crossings enters the result.
Phonopy applies degenerate-subspace perturbation theory when required;
explicit path-aligned checks at representative segment midpoints gave the same
projected velocities.  We retained Phonopy 2.43.4 for the analytical-derivative
evaluation and reproduced a representative case with the 4.1.0
finite-displacement runtime to numerical precision.  We
then define
\begin{equation}
 \mathcal{P}(T) \equiv \langle v^2\rangle_C =
 \frac{\sum_{qj} C_{qj}(T)\,[v_{qj}^{\parallel}]^{2}}
       {\sum_{qj} C_{qj}(T)},
 \label{eq:proxy}
\end{equation}
where the sum covers samples above 0.05 THz on the common band path.  We refer
to $\mathcal{P}$ as the band-path heat-capacity-weighted mean-square group
velocity.  Its square root is
the heat-capacity-weighted root-mean-square velocity, $v_{\rm rms}$.  The same
path, sampling, cutoff, and normalization were used for every twisted cell and
its matched controls.  Equation~(\ref{eq:proxy}) isolates a harmonic velocity
scale; it is not an absolute lattice thermal conductivity because it
does not contain full Brillouin-zone integration or anharmonic lifetimes.
Imaginary frequencies and positive frequencies below the stated cutoff do not
enter either the numerator or denominator.

The spectral decomposition used 0.10-THz bins at 300 K.  Within each bin we
computed the contribution to the numerator, normalized its frequency density
so that the retained-bin contributions integrate to unity, evaluated the ratio
of twisted and control $v_{\rm rms}$, and accumulated the heat-capacity share.
Temperature dependence
was evaluated at 100, 300, 500, and 700 K.  Figure~\ref{fig:phonon} summarizes
the 300-K angle dependence; the temperature test is reported in the text.

\subsection{Finite anisotropic crossover analysis}

Structural reconstruction was compared using the displacement from the stored
post-commensuration, pre-relaxation structure after removal of a common rigid
translation.
We report the root-mean-square (RMS) in-plane displacement.  To compare the
spatial texture across cells of different size, the in-plane displacement in
each layer was projected onto the lowest periodic Fourier harmonics
$|h|,|k|\leq1$ in reduced moir\'{e} coordinates after removal of the mean
translation of that layer.  The projected fields were evaluated on a common
$48\times48$ grid and compared by their Pearson correlation.

The approach to structural saturation was summarized by
\begin{equation}
 u_{\rm rms}(\theta)=
 \frac{u_\infty}{1+(\theta/\theta_{1/2})^m}.
 \label{eq:relax-fit}
\end{equation}
This empirical form tests the expected weak-relaxation power law and provides
a compact measure of saturation; $\theta_{1/2}$ is not assigned as a special
transport angle.  Spectral self-similarity was evaluated from the Pearson
correlation between adjacent-angle
$v_{\rm rms}^{\rm tw}/v_{\rm rms}^{\rm ctl}$ profiles in 56 common 0.10-THz
bins centered from 0.05 to 5.55 THz.

\subsection{DFT registry atlas}

Local bilayer registries were evaluated with VASP 6.1.0
\cite{Kresse1996} using the projector augmented-wave (PAW) method
\cite{Blochl1994}.  The registry calculations used PBE
\cite{PBE1996} with D3(BJ) dispersion \cite{Grimme2011},
\texttt{PAW\_PBE Sn\_d 06Sep2000} and
\texttt{PAW\_PBE Se 06Sep2000} data,
a 600-eV plane-wave cutoff, a $16\times16\times1$ $\Gamma$-centered mesh,
and an electronic tolerance of $10^{-8}$ eV.  Nine translations
$(i_x/3,i_y/3)$ with $i_x,i_y\in\{0,1,2\}$ were relaxed at fixed cell.  One
in-plane anchor in each layer prevented registry drift while all $z$
coordinates and the remaining in-plane coordinates were relaxed.  The maximum
force over free coordinates was below $10^{-3}$ eV \AA$^{-1}$ for every case,
and the final periodic vacuum gap exceeded 30 \AA.

Static convergence tests on four representative registries varied the cutoff
from 400 to 600 eV, the in-plane mesh from $8\times8$ to $16\times16$, and
the periodic vacuum gap by an additional 5 and 10 \AA.  The registry ordering was
unchanged in every test.  The maximum relative-energy shift from the
500-eV, $12\times12\times1$ reference was 0.37 meV per bilayer over all tests
and 0.11 meV per bilayer for the higher-accuracy cutoff, mesh, and vacuum
variants.

The six inequivalent registries were also calculated with the optB86b-vdW
functional \cite{Klimes2011} and the same Sn$_{\mathrm d}$/Se PAW data.  The minimum and
maximum were evaluated separately with standard-Sn PBE-D3(BJ) PAW data.  Only
relative energies within a fixed functional and PAW branch are compared.

\begin{figure*}[t!]
  \includegraphics[width=\textwidth]{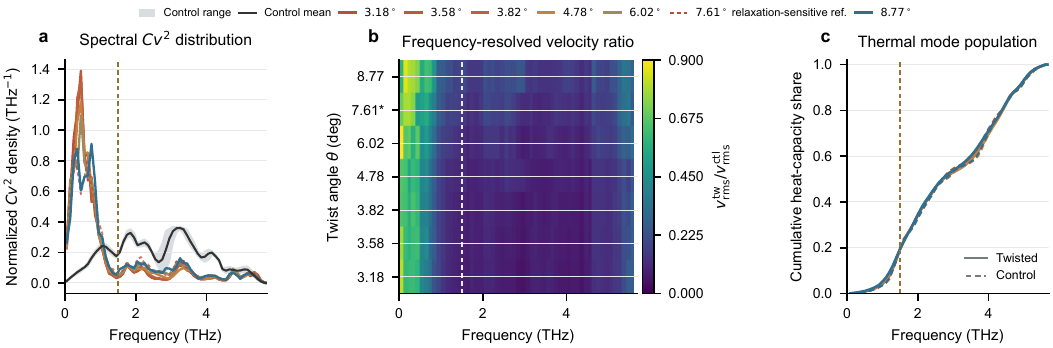}
  \caption{\label{fig:spectral}Frequency-resolved origin of the 300-K
  suppression.  (a) Normalized spectral $C v^2$ density for the calculated
  twisted cells; the relaxation-sensitive $7.61^\circ$ reference is dashed and
  the shaded band spans the seven per-angle matched-control means, while the
  black curve is their overall mean.  Each profile is divided
  by its retained $Cv^2$ sum and the bin width and therefore integrates to
  unity over frequency.  (b) Ratio of
  analytical twisted and per-angle control-mean $v_{\rm rms}$ in discrete
  0.10-THz bins.  Each row is one calculated angle and no angular
  interpolation is applied; the color scale spans the finite plotted values
  and the asterisk marks the relaxation-sensitive reference.
  (c) Cumulative harmonic heat-capacity share for $3.18^\circ$, $4.78^\circ$,
  and $8.77^\circ$ twisted cells (solid) and their per-angle control means
  (dashed).  The vertical dashed line in every panel marks 1.5 THz.}
\end{figure*}

\section{Results}
\label{sec:results}

\subsection{Relaxed vacuum-separated moir\'{e} bilayers}

Figure~\ref{fig:structure} shows all seven relaxed bilayers at the full
commensurate scale together with a representative side view.  The in-plane
cell lengths increase from $29.109\times26.949$ \AA{} at $8.77^\circ$ to
$79.454\times75.156$ \AA{} at $3.18^\circ$, while the atom count grows from
340 to 2588.  The explicit out-of-plane cell separates the bilayer from its
periodic image by approximately 25 \AA{} (Table~\ref{tab:cells}).

Relaxation produces a smooth variation of interlayer geometry across the
angle series.  The layer-center separation decreases from 5.904 \AA{} at
$8.77^\circ$ to 5.793 \AA{} at $3.18^\circ$
[Table~\ref{tab:cells}; Fig.~\ref{fig:structure}(h)].  The minimum three-dimensional interlayer
distance remains within 3.255--3.279 \AA{} for all seven cells.  These values
describe the separation of the puckered bilayers directly and are distinct
from the total out-of-plane lattice parameter, which also contains the vacuum.

\subsection{Angle-dependent harmonic velocity scale}

The complete phonon spectra extend to approximately 5.6 THz, and their
low-frequency minima are summarized in Table~\ref{tab:low-frequency}.
Throughout the figures and tables, imaginary modes are reported as negative
frequencies.  The six
structures used to define the six-structure equilibrium trend have band-path minima of
about $-0.02$ THz and no sampled mode below the $-0.05$-THz numerical
threshold.  Their uniform $4\times4\times1$ scans give minima between
$-0.0216$ and $-0.0139$ THz and likewise contain no mode below the threshold.
The separately marked $7.61^\circ$ reference initially exhibits one
$-0.0814$-THz feature near $S$, but the feature disappears after tighter
fixed-cell relaxation.  Its structural closure and endpoint spectrum are
reported in Appendix~\ref{app:displacement} and
Fig.~\ref{fig:appendix-mode}; this reference is not used in the equilibrium
trend.

The analytical 300-K $\mathcal{P}$ values are lower than the corresponding
matched-control means throughout the calculated series
[Fig.~\ref{fig:phonon}(b)].  Ordered from $8.77^\circ$ to $3.18^\circ$, the
300-K, 0.05-THz ratios are 0.0836, 0.0592, 0.0405, 0.0334, 0.0290, 0.0284,
and 0.0260; the second value belongs to the excluded $7.61^\circ$ reference.
Among the six cells without a resolved
instability at the stated threshold, the suppression strengthens on
approaching the finite anisotropic crossover and changes only weakly across its
smallest-angle side.  The same qualitative angular pattern is
retained from 100 to 700 K.

\subsection{Velocity-dominated spectral response}

The decomposition in Fig.~\ref{fig:spectral} identifies why
$\mathcal{P}$ decreases.  The spectral $C v^2$ weight is lower in the twisted
cells across nearly the full 0--5.6-THz interval
[Fig.~\ref{fig:spectral}(a)].  The suppression is therefore not associated
with one isolated optical branch.  The frequency-resolved $v_{\rm rms}$ ratios
show the same broad response [Fig.~\ref{fig:spectral}(b)], with the strongest
reduction at the smaller angles.

By contrast, the cumulative heat-capacity curves nearly overlap
[Fig.~\ref{fig:spectral}(c)].  The thermal population of the folded branches
therefore changes much less than the velocity-weighted numerator.  Within
Eq.~(\ref{eq:proxy}), the suppression is governed primarily by the reduced
velocity spectrum rather than by removal of thermally populated modes.  At
300 K, the mode heat capacities vary only weakly over most of the spectrum, so
the cumulative comparison is also close to a comparison of the corresponding
mode-count distributions.

\subsection{Finite anisotropic structural and spectral crossover}

The weak angle dependence on the small-angle side is defined relative to the
per-angle matched controls.  Both the twisted and control velocity scales vary
with the source-layer metrics and cell size; their ratio isolates the
additional renormalization associated with twist.  This integrated trend is
independently supported by the convergence of the frequency-resolved
velocity-ratio profiles at $3.82^\circ$, $3.58^\circ$, and $3.18^\circ$
[Fig.~\ref{fig:appendix-crossover}(b)].  The
relaxation field evolves in the same way.  Its raw RMS in-plane amplitude is
0.480, 0.493, and 0.515 \AA{} at the three smallest angles, corresponding to
adjacent increases of 2.8\% and 4.4\%; these percentages were calculated from
unrounded values.  After projection onto the lowest
periodic Fourier components in reduced coordinates, the adjacent textures have
correlations of 0.983 and 0.993, while their projected RMS-amplitude ratios are
1.026 and 1.035.  From $3.82^\circ$ to $3.18^\circ$,
$\mathcal P^{\rm tw}$ decreases from $17.060\times10^3$ to
$16.021\times10^3$ m$^2$ s$^{-2}$ (6.1\%), whereas the matched-control mean
increases from $587.607\times10^3$ to $617.107\times10^3$ m$^2$ s$^{-2}$
(5.0\%).  Across all three smallest angles,
$\mathcal P^{\rm tw}/\mathcal P^{\rm ctl}$ spans 0.02903--0.02596; the full
span divided by the three-angle mean is 11.1\%, while the same convention gives
5.6\% for the corresponding $v_{\rm rms}$ ratio.  The finite anisotropic
crossover is therefore identified by the joint approach to a self-similar
relaxed structure and a self-similar spectral renormalization
(Appendix~\ref{app:crossover}).

\subsection{Registry energetics as structural context}

\begin{figure}[t!]
  \centering
  \includegraphics[width=\columnwidth]{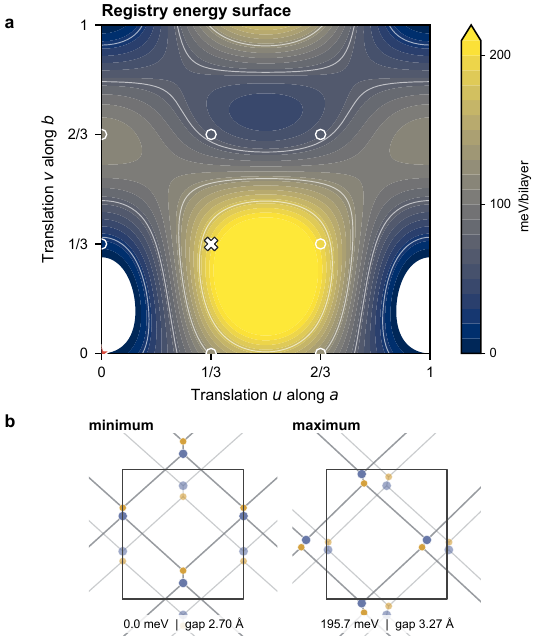}
  \caption{\label{fig:registry}DFT registry-energy landscape of bilayer SnSe.
  (a) Periodic Fourier guide through the $3\times3$ PBE-D3(BJ) values;
  circles are the DFT points, the star marks the minimum, and the cross marks
  the maximum.  Interpolation between points is a visual guide rather than
  additional calculated data.  (b) Matched top views of the relaxed minimum-
  and maximum-energy registries; the outlined rectangle is the primitive
  bilayer cell.}
\end{figure}

Figure~\ref{fig:registry}(a) summarizes the PBE-D3(BJ) registry-energy
landscape.
The nine calculated translations span 195.7 meV per bilayer, equivalent to
48.9 meV per SnSe unit.  The minimum occurs at $(i_x,i_y)=(0,0)$ and the
maximum at $(1,1)$.  The direct interlayer gap grows from 2.705 to 3.270 \AA{}
between these extrema [Fig.~\ref{fig:registry}(b)].  Across all nine cases, the
Pearson correlation between relative energy and direct gap is 0.998
(Appendix~\ref{app:registry}), showing that registry and relaxed separation are
tightly coupled in the primitive bilayer.

The registry ordering and corrugation scale remain consistent across the
tested dispersion treatments and Sn PAW choices
(Appendix~\ref{app:methods} and Table~\ref{tab:appendix-registry-methods}).
The landscape in Fig.~\ref{fig:registry} therefore provides a stable
structural reference for interpreting the moir\'{e} calculations.

The registry landscape supplies the energy scale of the local environments
sampled across a moir\'{e} cell.  The direct comparison in
Appendix~\ref{app:ifc} links these structures to their harmonic force response:
the Hessian components have $r=0.996$ and a normalized root-mean-square error
(NRMSE) of 0.091, while the
nine matched modes below 2 THz have a frequency MAE of 0.023 THz.  This
low-frequency agreement is the part of the local benchmark most directly
connected to the velocity descriptor.

\section{Discussion}
\label{sec:discussion}

The seven-angle data show that the heat-capacity-weighted mean-square
velocity is strongly reduced throughout the calculated range.  Among the six
structures without a resolved instability at the stated threshold, the
reduction strengthens toward the onset of the finite anisotropic crossover and then
becomes only weakly angle dependent.  This
response is not concentrated in one narrow frequency window.  It appears as a
broad reduction of
heat-capacity-weighted phonon velocities while the cumulative heat-capacity
distribution remains nearly unchanged.  Extensive folding alone increases the
number of branches in a commensurate description; the matched-control
construction shows that the observed reduction comes from the change in their
dispersion rather than from branch counting.

The structural and harmonic tests support this interpretation at two length
scales.  DFT gives a registry corrugation of 0.16--0.20 eV per bilayer and a
strong relation between local registry and relaxed interlayer separation.  The
calibrated MACE potential reproduces this ordering and agrees with the centered
DFT force response, Hessian, low-frequency $\Gamma$ modes, and $2\times2$
spatial decay.  In the full moir\'{e} cells, the same potential yields physical
interlayer distances after strict relaxation.  The six structures defining the
equilibrium trend have no path or $4\times4\times1$ q-grid modes below the
$-0.05$-THz numerical threshold.  The relaxation-sensitive $7.61^\circ$
reference loses its initial negative feature under tighter relaxation and does
not enter that trend.

A continuum description relates the finite anisotropic crossover to the competition
between intralayer elasticity and the registry-dependent interlayer energy.
Let $\bm u(\bm r)$ be the relative in-plane relaxation and
$\bm\delta_0(\bm r)=[\bm R(\theta)-\bm I]\bm r\simeq
\theta\hat{\bm z}\times\bm r$ be the unrelaxed local stacking field.  The
relaxed stacking field is $\bm\delta(\bm r)=\bm\delta_0(\bm r)+\bm u(\bm r)$.
A standard elastic--registry functional is
\begin{equation}
 \begin{aligned}
 E[\bm u]=\int_{\Omega_{\rm M}}d^2r\bigg[&
 \frac{1}{2}C^{\rm rel}_{ijkl}\partial_i u_j\partial_k u_l \\
 &+\sum_n V_n\cos\!\left(
 \bm G_n\!\cdot\bm\delta(\bm r)+\phi_n\right)\bigg],
 \end{aligned}
 \label{eq:continuum-energy}
\end{equation}
where $C^{\rm rel}_{ijkl}$ is the relative-layer elastic tensor and
$V(\bm\delta)=\sum_nV_n\cos(\bm G_n\!\cdot\bm\delta+\phi_n)$ is the stacking
energy per area.  This form follows established continuum descriptions of
relaxation and solitons in twisted bilayers
\cite{Nam2017,Naik2018,Carr2018,DeBeule2026}.  The phases $\phi_n$ allow for
the low symmetry of the puckered lattice.  The atomistically relaxed
interlayer separation and corrugation are absorbed into the effective
coefficients $V_n$ in this in-plane reduction.

Stationarity of Eq.~(\ref{eq:continuum-energy}) gives
\begin{equation}
 -C^{\rm rel}_{ijkl}\partial_i\partial_k u_l=
 \sum_n V_n G_{n,j}\sin\!\left(
 \bm G_n\!\cdot[\bm\delta_0+\bm u]+\phi_n\right).
 \label{eq:continuum-balance}
\end{equation}
In the weak-relaxation limit, the forcing associated with $\bm G_n$ varies
at the moir\'{e} wave vector
$\bm q_n\simeq\theta\hat{\bm z}\times\bm G_n$.  Linearizing
Eq.~(\ref{eq:continuum-balance}) therefore gives
$|\bm G_n\cdot\bm u_n|\sim\eta_n$, where
\begin{equation}
 \eta_n(\theta)=
 \frac{|\bm G_n|^2|V_n|}{\mathcal C_n|\bm q_n|^2}
 \simeq\frac{|V_n|}{\mathcal C_n\theta^2},
 \label{eq:relaxation-parameter}
\end{equation}
and $\mathcal C_n$ is the directional elastic modulus for the relevant
polarization.  This result supplies the $\theta^{-2}$ law directly from force
balance rather than treating it as an empirical angle dependence.

In the weak-relaxation regime, the reconstruction grows approximately as
$\theta^{-2}$.  Excluding the relaxation-sensitive $7.61^\circ$ reference,
Eq.~(\ref{eq:relax-fit}) gives $m=2.09$ for the six equilibrium structures.
The three-parameter fit is poorly constrained under leave-one-angle-out tests,
so neither this exponent nor the fitted half-saturation angle is interpreted
as a precision estimate.  The best-fit exponent is compatible with, but does
not independently establish, the $\theta^{-2}$ limit.  Once $\eta_n$ becomes
of order unity, favorable
registries expand and the mismatch is increasingly concentrated between them.
Balancing elastic and stacking terms gives the characteristic width
\begin{equation}
 w_n\sim\frac{1}{|\bm G_n|}
 \left(\frac{\mathcal C_n}{|V_n|}\right)^{1/2},
 \qquad
 \frac{L_{\rm M}}{w_n}\propto\sqrt{\eta_n},
 \label{eq:wall-scaling}
\end{equation}
so that $w_n$ is set mainly by material parameters while the moir\'{e} period
$L_{\rm M}$ continues to grow as $\theta^{-1}$
\cite{Nam2017,Naik2018,Carr2018,DeBeule2026}.  The relaxation texture consequently approaches a
fixed form in reduced moir\'{e} coordinates.  The observed exponent and the
convergence of the low-order relaxation harmonics are consistent with this
approach to the slowly varying small-angle regime
(Appendix~\ref{app:crossover}).  Equations
(\ref{eq:continuum-energy})--(\ref{eq:wall-scaling}) provide a scaling
interpretation; without independently fitted $C^{\rm rel}_{ijkl}$ and a
continuous stacking-energy surface, they are not a parameter-free prediction
of the crossover angle.

The same relaxed field enters the long-wavelength lattice dynamics.  For a
small relative-layer vibration $\bm\xi$ about the equilibrium texture
$\bm u_0$, the second variation contains a periodic registry stiffness
\begin{equation}
 \begin{aligned}
 \delta^2E=\frac{1}{2}\int d^2r\big[&
 C^{\rm rel}_{ijkl}\partial_i\xi_j\partial_k\xi_l \\
 &+\xi_i K_{ij}(\bm r)\xi_j\big], \\
 K_{ij}(\bm r)={}&
 \left.\partial_{\delta_i}\partial_{\delta_j}V(\bm\delta)
 \right|_{\bm\delta=\bm\delta_0+\bm u_0}.
 \end{aligned}
 \label{eq:continuum-dynamics}
\end{equation}
In a moir\'{e} reciprocal basis, this stiffness couples components separated by
moir\'{e} reciprocal vectors according to
\begin{equation}
 D_{ij}(\bm k+\bm g,\bm k+\bm g')=
 D^{\rm el}_{ij}(\bm k+\bm g)\delta_{\bm g\bm g'}+
 \rho^{-1}K_{ij}(\bm g-\bm g').
 \label{eq:moire-dynamical-matrix}
\end{equation}
Here $\rho$ is the reduced areal mass density for relative-layer motion, and
$K_{ij}(\bm g)$ denotes the Fourier component of the Hessian of the same
stacking-energy density $V(\bm\delta)$ defined below
Eq.~(\ref{eq:continuum-energy}).
Thus a self-similar stacking texture produces a self-similar distribution of
registry-induced mode coupling.  This provides a dynamical connection between
the structural convergence and the nearly collapsed frequency-resolved
velocity ratios in Fig.~\ref{fig:appendix-crossover}.  Equations
(\ref{eq:continuum-dynamics}) and (\ref{eq:moire-dynamical-matrix}) are a
long-wavelength description rather than a replacement for the atomistic
spectrum; the puckered lattice mixes in-plane and out-of-plane polarizations,
which are retained in the centered finite-displacement calculations.  The limiting
behavior is consistent with continuum descriptions of relaxed moir\'{e}
phonons and small-angle pinned-domain dynamics
\cite{Koshino2019,Lu2022,Ochoa2019,Maity2020,Gadelha2021,RamosAlonso2025}.

The atomistic descriptors place the observed crossover between
$4.78^\circ$ and $3.82^\circ$.  Because the elastic moduli and stacking
harmonics of rectangular SnSe are direction dependent, the leading $\eta_n$
are not symmetry equivalent and need not reach the nonlinear regime at one
angle.  The finite, broadband response is therefore more naturally described
by a finite anisotropic crossover than by a singular angle.  The smaller-angle
cells sample the slowly varying side of this interval.

The connection to thermoelectric transport follows from the modal form
\begin{equation}
 \kappa_{\alpha\beta}=\frac{1}{\mathcal V}
 \sum_{\bm qj}C_{\bm qj}v_{\bm qj,\alpha}
 v_{\bm qj,\beta}\tau_{\bm qj}.
 \label{eq:kappa-modal}
\end{equation}
The present $\mathcal P=\langle v^2\rangle_C$ isolates the harmonic velocity
factor in Eq.~(\ref{eq:kappa-modal}).  The nearly unchanged cumulative heat
capacity and the broad reduction of the velocity spectrum identify this factor
as the principal source of the calculated suppression.  Angles near the onset
of the crossover may therefore retain much of the harmonic velocity reduction
without the rapid increase in moir\'{e} cell size at still smaller angles.

The descriptor does not include the full Brillouin zone or three- and
four-phonon scattering, which can be important in lone-pair chalcogenides
\cite{Wan2025BiCuSeO,Bai2025SrSnSe2}, and it should therefore not be read as an
absolute lattice thermal conductivity.  The short selected-mode dynamics in
Appendix~\ref{app:lifetime} is retained only as a diagnostic and is not
combined with the equilibrium harmonic descriptor.  A complete transport
calculation will require mode-resolved anharmonic lifetimes over the full
Brillouin zone and electronic transport at stable relaxed structures.

Within this scope, the result identifies a clear physical role of twist in
puckered SnSe.  The registry pattern reconstructs and lowers the
heat-capacity-weighted harmonic velocity scale over a broad frequency range.
As the angle decreases, the elastic--registry
parameter in Eq.~(\ref{eq:relaxation-parameter}) grows until the relaxed
stacking texture and the normalized velocity spectrum approach a self-similar
form.  The anisotropy of SnSe spreads this change over a finite angular
interval, and the resulting crossover is supported by both structural and
spectral observables.

\section{Conclusions}
\label{sec:conclusions}

We have established broadband suppression of harmonic phonon velocities and a
finite anisotropic crossover in twisted, vacuum-separated bilayer SnSe using
seven relaxed commensurate cells containing 340--2588 atoms.  Centered
finite-displacement calculations and per-angle matched controls show a
reduction of $\langle v^2\rangle_C$ to 2.6--8.4\% of the matched-control
values at 300 K, and a weakly angle-dependent response on the small-angle
side of the crossover, spanning 11.1\% of its mean across the three smallest
angles.  Uniform q-grid scans and tighter structural checks establish the
six-structure equilibrium trend within the quantified low-frequency
resolution.  The thermally occupied mode population changes
little, whereas the group velocities are reduced throughout the spectrum.  DFT registry energetics and a direct DFT--MACE
force-constant comparison support the local structural and harmonic
description used for these large cells.

The elastic--registry framework organizes these observations through
reconstruction parameters $\eta_n\propto\theta^{-2}$.  As they enter the
nonlinear regime, the relaxed stacking texture and its registry-induced
dynamical coupling approach self-similar forms.  Because the directional
elastic moduli and stacking harmonics of rectangular SnSe are not equivalent,
this change is spread over a finite angular interval, identified here between
$4.78^\circ$ and $3.82^\circ$.  The simultaneous
convergence of the reduced-coordinate relaxation texture and the
frequency-resolved velocity ratio is consistent with this finite anisotropic
crossover and explains the slowly varying response at the smallest
calculated angles.

These results place interlayer twist alongside composition, strain, and defect
engineering as a structural variable for controlling phonons in SnSe.  Angles
near the onset of the crossover may offer a useful compromise: they retain much
of the harmonic velocity suppression while avoiding the largest moir\'{e}
cells.  Establishing the resulting thermoelectric benefit requires full
Brillouin-zone anharmonic lifetimes and electronic transport at the same
relaxed geometries, so that the changes in lattice thermal conductivity,
Seebeck coefficient, and electrical conductivity can be assessed together.

A registry-resolved configuration-space dynamical theory would extend the
present discrete angle series to continuously variable twist, pressure, and
strain.  Such a framework could determine how directional elasticity and
stacking harmonics broaden or narrow the crossover and could identify which
phonon polarizations are most responsive to external control.  The present
structural and harmonic results provide a basis for that development and for
full thermoelectric optimization of twist-controlled layered materials.

\section*{Data availability}

The data that support the findings of this study are available from the
corresponding authors upon reasonable request.

\section*{Acknowledgments}

This work was supported by the National Science Fund for Distinguished Young
Scholars (No. 51925101), the National Natural Science Foundation of China
(Nos. 52450001 and 12104370), the Tianmushan Laboratory Research Project
(Nos. TK2024D006 and TK2023C021), the Beijing Natural Science Foundation
(No. JQ18004), the 111 Project (No. B17002), and the Tencent Xplorer Prize.
Additional support was provided by the Academic Excellence Foundation of BUAA
for PhD Students.  The authors acknowledge the high-performance computing
resources provided by Tianmushan Laboratory and Beihang University.

\appendix

\setcounter{figure}{0}
\renewcommand{\thefigure}{A\arabic{figure}}
\renewcommand{\theHfigure}{A.\arabic{figure}}
\setcounter{table}{0}
\renewcommand{\thetable}{A\arabic{table}}
\renewcommand{\theHtable}{A.\arabic{table}}

\begin{figure}[t!]
  \centering
  \includegraphics[width=\columnwidth]{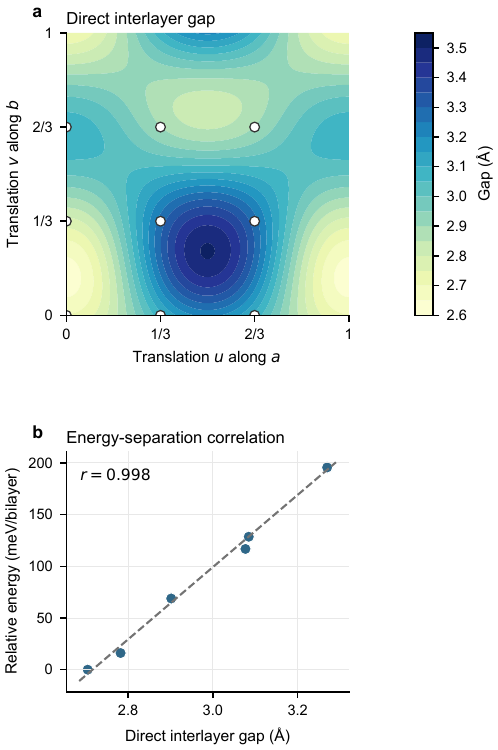}
  \caption{\label{fig:appendix-gap}Registry geometry diagnostics.  (a) Periodic
  Fourier guide through the direct interlayer gaps of the nine PBE-D3(BJ)
  registries.  Circles mark calculated translations.  (b) Within-branch
  relative energy versus direct interlayer gap; the dashed line is a
  least-squares guide.}
\end{figure}

\begin{figure*}[t!]
  \includegraphics[width=\textwidth]{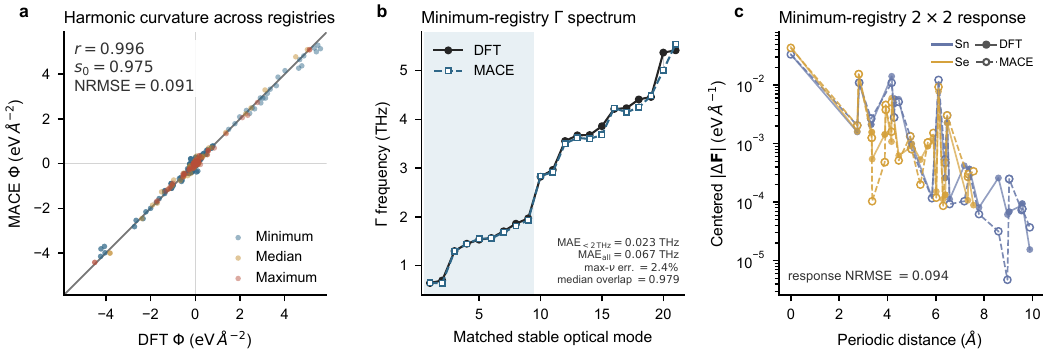}
  \caption{\label{fig:ifc-benchmark}Direct local MACE--DFT harmonic benchmark.
  (a) MACE versus DFT force-constant components for the full minimum-registry
  primitive-cell Hessian and selected median- and maximum-registry columns; the
  diagonal is equality.  (b) Matched stable optical $\Gamma$ modes after the
  same symmetrization and minimal acoustic-sum projection is applied to both
  minimum-registry Hessians; the blue band spans the nine matched modes below
  2 THz.  (c) Centered $2\times2$ force-response magnitude
  versus periodic distance for representative Sn and Se displacements.  Solid
  filled traces are DFT and dashed open traces are MACE.}
\end{figure*}

\section{Commensurate cells and full phonon calculations}
\label{app:construction}

The rectangular commensurate family uses
\begin{equation}
 \bm{M}_{\mathrm{L}}=\begin{pmatrix}p&0\\0&q\end{pmatrix},\qquad
 \bm{M}_{\mathrm{U}}=\begin{pmatrix}p&s\\-t&q\end{pmatrix},
\end{equation}
which repeats Eq.~(\ref{eq:commensurate-matrices}) for convenience.  With
$s=t=1$, the lower and upper
determinants are $pq$ and $pq+1$,
respectively.  Since a monolayer primitive cell contains four atoms, the
twisted bilayer contains $4(2pq+1)$ atoms.  Table~\ref{tab:cells} lists the
seven cells after relaxation.  The symmetric area strain is below 0.60\% and
the aspect-ratio strain is below 0.81\% throughout the series.

The two four-atom Sn$_2$Se$_2$ source motifs were obtained from the DFT-relaxed
layer orientations used for the registry atlas; their in-plane metrics are
listed with the relaxed moir\'e cells in Table~\ref{tab:cells}.  Full fractional
coordinates are not repeated here; the corresponding POSCAR and extended-XYZ
files are available from the corresponding authors upon reasonable request and
preserve the exact atom ordering, cell vectors, and construction origin.
Exact closure was imposed by equal and opposite logarithmic area accommodation,
so neither layer carries the full commensuration strain.  The
primitive-index-zero Sn sites define the common rotation center at fractional
coordinate $(1/2,1/2)$ and the stored pre-relaxation reference.  The untwisted
controls use the registry-minimum translation $(u,v)=(1/3,0)$ in the
corresponding source basis.

\begin{table}[h!]
\caption{\label{tab:cells}Source-layer and relaxed commensurate-cell metrics.
For the moir\'e cells, $N$ is the atom count and the last column lists
$d_{\rm cent}$ and the periodic vacuum gap.  Lengths are in \AA.}
\centering
\begin{ruledtabular}
\begin{tabular}{lrrr}
\multicolumn{4}{c}{Source monolayers} \\
Source & $a$ & $b$ & $\gamma$ (deg) \\
\hline
Lower & 4.17229 & 4.46353 & 90.000 \\
Upper & 4.13558 & 4.42415 & 90.011 \\
\hline
\multicolumn{4}{c}{Relaxed commensurate bilayers} \\
$\theta$ (deg) & $(p,q);N$ & $a\times b$ & $d_{\rm cent};d_{\rm vac}$ \\
\hline
8.77 & $(7,6);340$   & $29.109\times26.949$ & $5.904;24.933$ \\
7.61 & $(8,7);452$   & $33.390\times31.234$ & $5.879;24.864$ \\
6.02 & $(10,9);724$  & $41.963\times39.809$ & $5.847;24.844$ \\
4.78 & $(13,11);1148$ & $53.662\times49.362$ & $5.825;24.665$ \\
3.82 & $(16,14);1796$ & $66.560\times62.261$ & $5.805;24.647$ \\
3.58 & $(17,15);2044$ & $70.858\times66.560$ & $5.801;24.770$ \\
3.18 & $(19,17);2588$ & $79.454\times75.156$ & $5.793;24.785$ \\
\end{tabular}
\end{ruledtabular}
\end{table}

Positive and negative displacements along all Cartesian directions require
$6N$ force evaluations without symmetry reduction.  Table~\ref{tab:force}
summarizes the full calculations.  The lower and upper controls contain
$8pq$ and $8(pq+1)$ atoms, respectively.

\begin{table*}[t!]
\caption{\label{tab:force}Centered finite-displacement phonons.  Control values
are listed as lower; upper.  The $7.61^\circ$ row is the original
relaxation-sensitive reference; its tight-relaxation closure is reported in
Appendix~\ref{app:displacement}.}
\begin{ruledtabular}
\begin{tabular}{lrrrrr}
$\theta$ (deg) & Twist atoms & Twist forces & Control forces &
$\nu_{\min}^{\rm tw}$ (THz) & $\nu_{\min}^{\rm ctl}$ (THz) \\
\hline
8.77 & 340  & 2040  & 2016; 2064   & $-0.0220$ & $-0.0094$; $-0.0223$ \\
7.61 & 452  & 2712  & 2688; 2736   & $-0.0814$ & $-0.0086$; $-0.0199$ \\
6.02 & 724  & 4344  & 4320; 4368   & $-0.0163$ & $-0.0116$; $-0.0184$ \\
4.78 & 1148 & 6888  & 6864; 6912   & $-0.0206$ & $-0.0130$; $-0.0175$ \\
3.82 & 1796 & 10776 & 10752; 10800 & $-0.0169$ & $-0.0135$; $-0.0163$ \\
3.58 & 2044 & 12264 & 12240; 12288 & $-0.0163$ & $-0.0138$; $-0.0163$ \\
3.18 & 2588 & 15528 & 15504; 15552 & $-0.0148$ & $-0.0142$; $-0.0161$ \\
\end{tabular}
\end{ruledtabular}
\end{table*}

\section{DFT registry energetics and geometry}
\label{app:registry}

The registry calculations used PBE-D3(BJ), Sn$_{\mathrm d}$/Se PAW data, 600 eV,
$16\times16\times1$ sampling, $10^{-8}$-eV electronic convergence, and a
$10^{-3}$-eV-\AA$^{-1}$ force threshold over free coordinates.  The relaxed
structures preserve the bilayer stoichiometry and a periodic vacuum gap
exceeding 30 \AA.

The nine registries span direct interlayer gaps of 2.705--3.270 \AA{}
and layer-center separations of 5.721--6.124 \AA.  Relative energy has
Pearson correlations of 0.998 with the direct gap and 0.987 with the
layer-center separation.  Figure~\ref{fig:appendix-gap} shows the direct-gap
landscape and its correlation with registry energy.  The interpolation is a
periodic Fourier guide constrained by the $3\times3$ points.

\begin{figure*}[t!]
  \centering
  \includegraphics[width=0.92\textwidth]{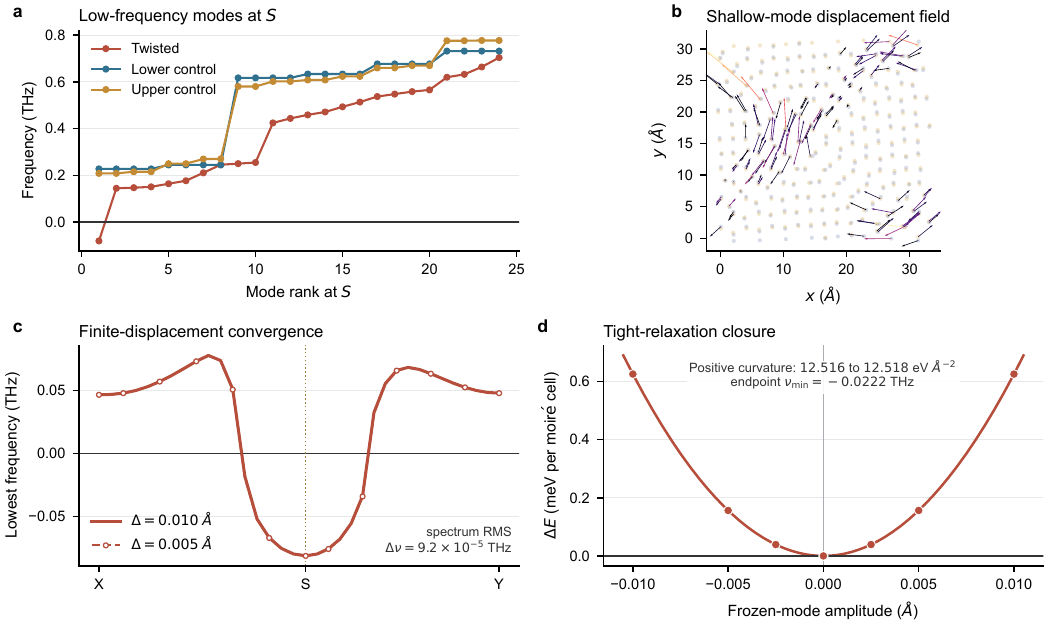}
  \caption{\label{fig:appendix-mode}Relaxation sensitivity of the shallow
  $7.61^\circ$ mode.  (a) Lowest 24 frequencies at $S$ for the twisted
  cell and its matched controls.  (b) In-plane displacement field of the shallow
  twisted mode.  (c) Lowest branch between $X$ and $Y$ for 0.005- and
  0.010-\AA{} centered displacements at the original structure.  (d) Centered
  energy probes along the same eigenvector after tighter fixed-cell
  relaxation; circles are MACE energies and the solid line is the mean
  quadratic curvature.  The annotation also gives the minimum of the centered
  1808-atom endpoint spectrum.}
\end{figure*}

\section{Direct MACE--DFT comparison of harmonic force constants}
\label{app:ifc}

The harmonic comparison used identical cells, coordinates, atom ordering, and
centered $\pm0.01$-\AA{} displacements for VASP and the calibrated MACE model.
The DFT settings were PBE-D3(BJ), Sn$_{\mathrm d}$/Se PAWs, 600 eV,
$16\times16\times1$ sampling for the primitive cells,
$8\times8\times1$ for the $2\times2$ cell, \texttt{LREAL=.FALSE.}, and
\texttt{ISYM=0}, with an electronic tolerance of $10^{-8}$ eV.  Forces were
evaluated for four reference structures and 76 displaced structures using the
same geometries and atom ordering in both methods.

The primitive-cell set contains all 24 columns of the minimum-registry
$24\times24$ Hessian and six selected columns spanning Sn/Se and $x/y/z$ at
each of the median and maximum registries, for 36 columns in total.  Each
centered odd force response has 24 components, giving 864 independent Hessian
components.  The 1728 raw positive/negative response components are the paired
one-sided samples from which those centered components are formed, rather than
a second independent benchmark.  The $2\times2$ set probes representative Sn
and Se $x$ columns and resolves their spatial force-response decay.

The 864 centered Hessian components give Pearson $r=0.996$, an
origin-constrained slope of 0.975, and an NRMSE of 0.091
[Fig.~\ref{fig:ifc-benchmark}(a)].  The slope corresponds to an average MACE
curvature 2.5\% below DFT under the constrained fit.  This small common bias is
well below the full component scatter and does not change the registry-resolved
signs or ordering; matched-control ratios reduce common scale errors but are
not assumed to remove them exactly.  The largest normalized Hessian error among
the registry, species, and direction subsets is 0.109.

The same symmetrization and acoustic-sum projection was applied to both
minimum-registry Hessians.  The 21 stable optical modes have a 0.067-THz
frequency MAE, 2.37\% maximum-frequency error, and median mass-weighted overlap
0.979 [Fig.~\ref{fig:ifc-benchmark}(b)].  The $2\times2$ centered response has
Pearson $r=0.996$, slope 0.977, and NRMSE 0.094, and it follows the DFT
spatial decay across the periodic cell [Fig.~\ref{fig:ifc-benchmark}(c)].  All
Hessians were obtained from the centered odd force response of each method,
which isolates the local harmonic curvature at the matched structures.

\section{Shallow mode and finite-displacement convergence}
\label{app:displacement}

The centered $7.61^\circ$ spectra calculated with 0.005- and 0.010-\AA{}
displacements have a root-mean-square difference of
$9.24\times10^{-5}$ THz and a maximum branch difference of
$8.19\times10^{-4}$ THz.  Their minimum frequencies differ by only
$1.47\times10^{-4}$ THz.  The shallow mode has participation ratio 0.506,
92.8\% in-plane weight, and lower/upper layer weights of 54.2\%/45.8\%.
Tightening the 452-atom fixed-cell relaxation from the common $10^{-3}$ to
$10^{-5}$ eV \AA$^{-1}$ threshold reaches
$f_{\max}=9.98097\times10^{-6}$ eV \AA$^{-1}$, lowers the energy by only
2.71 meV per moir\'{e} cell, and produces an RMS atomic displacement of
0.103 \AA.  The displacement is periodic in the original cell and therefore
belongs to the $\Gamma$ sector; after embedding it in a $2\times2$ cell, its
mass-weighted overlap with the original $S=(1/2,1/2,0)$ Bloch eigenvector is
$1.9\times10^{-18}$.  Thus the tighter relaxation changes the $S$-point
curvature without condensing that mode.  At the
tightly relaxed structure, centered probes of the released $S$-point
eigenvector at 0.0025, 0.005, and 0.010 \AA{} give positive energy curvatures
of 12.5163, 12.5166, and 12.5178 eV \AA$^{-2}$ per moir\'{e} cell; the
independent force-projected curvatures are also positive.

For the endpoint diagnostic, the original 452-atom cell was expanded
$2\times2\times1$, which folds $S$ to $\Gamma$, seeded by 0.12 \AA{} along the
accepted $S$ eigenvector, and relaxed at fixed cell to the same
$10^{-5}$-eV-\AA$^{-1}$ criterion.  The resulting 1808-atom structure is lower
than the original reference by 0.42562 eV per 452-atom moir\'{e} cell and has a
1.4075-\AA{} projection on the initial folded mode; endpoints from opposite
seeds differ by only $7.2\times10^{-8}$ eV per original cell.  Its centered
phonon spectrum gives $\nu_{\min}=-0.02215$ THz, no path samples below
$-0.05$ THz, and a maximum $\Gamma$ acoustic residual of
$2.21\times10^{-7}$ THz.  With the same analytical-derivative convention,
its 300-K, 0.05-THz descriptors are
$\mathcal{P}=2.2064\times10^{4}$ m$^{2}$ s$^{-2}$ and
$v_{\rm rms}=148.54$ m s$^{-1}$.  This endpoint documents the original folded
landscape, but the pronounced negative feature does not survive tighter
relaxation of the primitive moir\'{e} cell.  The original point is therefore
retained only as the relaxation-sensitive $7.61^\circ$ reference and is not
used to define the six-structure equilibrium trend.

\begin{figure*}[t!]
  \includegraphics[width=\textwidth]{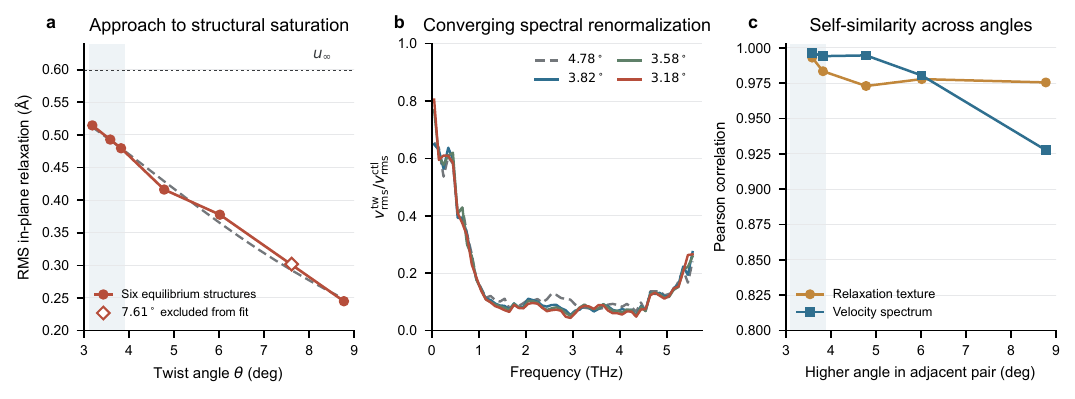}
  \caption{\label{fig:appendix-crossover}Structural and spectral signatures of
  the finite anisotropic crossover.  (a) RMS in-plane relaxation relative to the
  unrelaxed commensurate cells.  The dashed curve is the six-equilibrium-point
  fit to Eq.~(\ref{eq:relax-fit}), the horizontal line marks its fitted
  asymptote, and the open diamond is the excluded relaxation-sensitive
  $7.61^\circ$ reference.  (b) Frequency-resolved twisted-to-control
  $v_{\rm rms}$ ratios
  around the crossover.  (c) Adjacent-angle correlations of the lowest-order
  relaxation texture and the frequency-resolved velocity-ratio profile.  Each
  pair is plotted at its higher-angle member; the rightmost point therefore
  represents the $8.77^\circ$--$6.02^\circ$ equilibrium pair, with the
  relaxation-sensitive $7.61^\circ$ reference excluded.  The pale region in
  (a) and (c) marks $3.18^\circ$--$3.82^\circ$.  The restricted vertical ranges
  in (a) and (c) resolve changes within the calculated series.}
\end{figure*}

\section{Selected-mode dynamics}
\label{app:lifetime}

Selected-mode molecular dynamics used ASE 3.27.0 and MACE 0.3.16 for the
original $7.61^\circ$ reference and its two controls.  For each of two seeds,
2.048 ps of 300-K Langevin thermalization with friction 0.01 fs$^{-1}$ was
followed by velocity rescaling and a 20-ps microcanonical (NVE) trajectory with a
1-fs step.  Mass-weighted velocities were projected onto 64 $\Gamma$ modes in
the 0.5--2.5-THz window.  After mean subtraction and Hann windowing, direct
half-height interpolation gave the spectral full width at half maximum and
$\tau=1/(\pi\,\mathrm{FWHM})$.  The 20-ps record gives 0.05-THz frequency
spacing, so this test is qualitative rather than an equilibrium lifetime
calculation.  All six trajectories passed the temperature, energy-drift, and
linewidth-coverage criteria.

The two-seed set of 20-ps trajectories (Table~\ref{tab:modal-md}) gives
twisted-to-control median lifetime ratios of 0.941 and 0.989 in the selected
0.5--2.5-THz $\Gamma$-mode window.

\begin{table}[h!]
\caption{\label{tab:modal-md}Selected $\Gamma$-mode dynamics for the original
$7.61^\circ$ reference and its controls.  The lifetime proxy $\tau$ is reported
in ps as the median over 64 projected $\Gamma$ modes, and
$\bar\tau_{\rm ctl}$ is the mean of the two controls.  Values are diagnostic
only.}
\centering
\begin{ruledtabular}
\begin{tabular}{lrrrr}
Seed & $\tau_{\rm tw}$ & $\tau_{\rm L}$ & $\tau_{\rm U}$ &
$\tau_{\rm tw}/\bar\tau_{\rm ctl}$ \\
\hline
1 & 3.187 & 3.270 & 3.505 & 0.941 \\
2 & 3.287 & 3.335 & 3.309 & 0.989 \\
\end{tabular}
\end{ruledtabular}
\end{table}

The six trajectories have mean temperatures of 291.9--307.6 K and absolute
NVE energy drifts below $3.7\times10^{-7}$ eV atom$^{-1}$.

\section{Structural and spectral signatures of the finite anisotropic crossover}
\label{app:crossover}

The RMS in-plane relaxation increases from 0.245 \AA{} at $8.77^\circ$ to
0.515 \AA{} at $3.18^\circ$ [Fig.~\ref{fig:appendix-crossover}(a)].  Excluding
the relaxation-sensitive $7.61^\circ$ reference, a fit to the six equilibrium
structures in Eq.~(\ref{eq:relax-fit}) gives $u_\infty=0.599$ \AA{},
$\theta_{1/2}=7.45^\circ$, and $m=2.09$, with an RMS residual of 0.0076 \AA.
The leave-one-angle-out ranges are broad: 0.562--1.464 \AA{},
$1.40^\circ$--$7.94^\circ$, and 0.73--2.51, respectively.  The fitted curve is
therefore used only as a descriptive saturation guide.  Its best-fit exponent
is compatible with the $\theta^{-2}$ weak-relaxation scaling in
Eq.~(\ref{eq:relaxation-parameter}), but the individual fit parameters are not
precision determinations.

The lowest-order reduced-coordinate relaxation textures become more similar as
the angle decreases.  Their adjacent-angle correlations are 0.9835 for
$3.82^\circ/3.58^\circ$ and 0.9931 for
$3.58^\circ/3.18^\circ$.  The corresponding RMS texture-amplitude ratios are
1.026 and 1.035.  The analytical frequency-resolved velocity-ratio profiles
also approach a common small-angle form
[Fig.~\ref{fig:appendix-crossover}(b)].  These independent structural and
harmonic measures locate the crossover between the $4.78^\circ$ and
$3.82^\circ$ cells.  We do not interpret residual differences of a few percent
among the smallest angles as resolved fine structure.

\section{Low-frequency spectrum and stability overview}
\label{app:low-frequency}

Table~\ref{tab:low-frequency} compares the low-frequency minima of all seven
relaxed twisted cells and their controls along the same
$\Gamma$--$X$--$S$--$Y$--$\Gamma$ path and reports the independent
$4\times4\times1$ q-grid minima.  Six cells have band-path minima from
$-0.0220$ to $-0.0148$ THz and no path samples below $-0.05$ THz.  The
$7.61^\circ$ cell is the only exception, with an isolated minimum of
$-0.0814$ THz near $S$; its lower and upper matched controls have
minima of $-0.0086$ and $-0.0199$ THz, respectively.

\begin{table}[h!]
\caption{\label{tab:low-frequency}Minimum band-path frequencies of the twisted
cells and matched controls together with uniform $4\times4\times1$
q-grid minima for the twisted cells.  Control values are listed as lower; upper.
$N_{<-0.05}^{\rm grid}$ is the number of q-grid modes below $-0.05$ THz;
negative values denote imaginary frequencies.}
\begin{ruledtabular}
\begin{tabular}{lrrr}
$\theta$ (deg) & $\nu_{\min}^{\rm tw}$ / $\nu_{\min}^{\rm ctl}$ (THz) &
$\nu_{\min}^{\rm grid}$ (THz) & $N_{<-0.05}^{\rm grid}$ \\
\hline
8.77 & $-0.0220$ / $-0.0094$; $-0.0223$ & $-0.0216$ & 0 \\
7.61 & $-0.0814$ / $-0.0086$; $-0.0199$ & $-0.0814$ & 1 \\
6.02 & $-0.0163$ / $-0.0116$; $-0.0184$ & $-0.0164$ & 0 \\
4.78 & $-0.0206$ / $-0.0130$; $-0.0175$ & $-0.0181$ & 0 \\
3.82 & $-0.0169$ / $-0.0135$; $-0.0163$ & $-0.0142$ & 0 \\
3.58 & $-0.0163$ / $-0.0138$; $-0.0163$ & $-0.0144$ & 0 \\
3.18 & $-0.0148$ / $-0.0142$; $-0.0161$ & $-0.0139$ & 0 \\
\end{tabular}
\end{ruledtabular}
\end{table}

The character and displacement-amplitude convergence of the original
$7.61^\circ$ branch are reported once in
Appendix~\ref{app:displacement}.  Although its original frequency lies beyond
the approximately 0.02-THz floor defined by the other cells and controls, the
feature disappears after tighter relaxation and is not assigned as an
intrinsic instability.  The low-frequency spectral redistribution otherwise
remains common across the angle series.
A representative relaxation-threshold audit was also performed for the stable
$8.77^\circ$ twisted cell.  Tightening its fixed-cell relaxation from
$10^{-3}$ to $10^{-5}$ eV \AA$^{-1}$ reached
$f_{\max}=9.9569\times10^{-6}$ eV \AA$^{-1}$ and lowered the energy by
4.51 meV per 340-atom cell.  The raw displacement was 0.0865 \AA{} RMS,
primarily an opposing rigid translation of the two layers; removing those
layer translations leaves an internal RMS displacement of 0.0404 \AA.
Centered probes of the accepted $S$-point mode at 0.005 and 0.010 \AA{} retain
positive energy curvature, 20.916 and 20.917 eV \AA$^{-2}$ per moir\'{e} cell,
respectively.  This independent check supports the stable sign classification
of the $8.77^\circ$ reference under a two-orders-of-magnitude tighter force
criterion.
For the $8.77^\circ$ triplet, replacing the preserved pre-ASR force constants
with the production force constants changes $\mathcal P$ by less than
0.0003\% and $v_{\rm rms}$ by less than 0.00013\%.  The translational row-sum
residual of the twisted cell decreases from
$3.16\times10^{-4}$ to $1.24\times10^{-14}$ eV \AA$^{-2}$.  At 300 K, the
$8.77^\circ$ $\mathcal P$ ratio is 0.08511, 0.08364, and 0.08163 for cutoffs
of 0.02, 0.05, and 0.10 THz, respectively.  The percentage changes below were
evaluated from unrounded values.  Doubling the analytical path density from 9
to 18 points per segment changes the $\mathcal P$ ratio by
$-1.80$\% at $8.77^\circ$ and $-1.17$\% at $4.78^\circ$; the corresponding
$v_{\rm rms}$-ratio changes are $-0.90$\% and $-0.59$\%.  These checks are
small compared with the angle-dependent suppression.  The absolute values in
Table~\ref{tab:path-density} show that the control scales are nearly unchanged
by path refinement and that the small ratio shifts arise mainly from the
twisted spectra.

\begin{table}[h!]
\caption{\label{tab:path-density}Analytical path-density sensitivity at 300 K
and a 0.05-THz cutoff.  $n_q$ is the number of midpoints per path segment;
$\mathcal P^{\rm tw}$ and $\mathcal P^{\rm ctl}$ are in
$10^3$ m$^2$ s$^{-2}$.}
\centering
\begin{ruledtabular}
\begin{tabular}{lrrrr}
$\theta$ (deg) & $n_q$ & $\mathcal P^{\rm tw}$ &
$\mathcal P^{\rm ctl}$ & Ratio \\
\hline
8.77 & 9  & 48.827 & 583.742 & 0.08364 \\
8.77 & 18 & 47.937 & 583.617 & 0.08214 \\
4.78 & 9  & 18.396 & 550.013 & 0.03345 \\
4.78 & 18 & 18.177 & 549.938 & 0.03305 \\
\end{tabular}
\end{ruledtabular}
\end{table}

\section{Registry-energy consistency across DFT descriptions}
\label{app:methods}

The six inequivalent optB86b-vdW/Sn$_{\mathrm d}$ cases reproduce the
PBE-D3(BJ)/Sn$_{\mathrm d}$
ordering and extrema [Table~\ref{tab:appendix-registry-methods}], with
corrugations of 163.330 and 195.651 meV per bilayer, respectively.  A
PBE-D3(BJ) branch using standard Sn instead of Sn$_{\mathrm d}$ gives
198.065 meV per
bilayer between the same extrema.  Thus the registry ordering and corrugation
scale are consistent across the tested dispersion treatments and Sn PAW
choices.  Relative energies are compared only within each functional and PAW
branch.

\begin{table}[b!]
\caption{\label{tab:appendix-registry-methods}Registry-energy consistency
across DFT descriptions.  Entries are relative energies in meV per bilayer
within each method.  D3-$d$, vdW-$d$, and D3-Sn denote
PBE-D3(BJ)/Sn$_{\mathrm d}$, optB86b-vdW/Sn$_{\mathrm d}$, and
PBE-D3(BJ)/standard Sn, respectively.  A dash denotes
a registry not evaluated in the two-point standard-Sn check.}
\begin{ruledtabular}
\begin{tabular}{lrrr}
$(i_x,i_y)$ & D3-$d$ & vdW-$d$ & D3-Sn \\
\hline
$(0,0)$ &   0.000 &   0.000 &   0.000 \\
$(0,1)$ &  16.096 &   5.923 &       -- \\
$(1,2)$ &  68.860 &  51.628 &       -- \\
$(0,2)$ & 116.796 &  93.225 &       -- \\
$(1,0)$ & 128.600 & 102.034 &       -- \\
$(1,1)$ & 195.651 & 163.330 & 198.065 \\
\end{tabular}
\end{ruledtabular}
\end{table}

\bibliography{references}

\end{document}